\documentclass[a4paper,12pt]{article}
\usepackage[utf8]{in
    putenc}
\usepackage{cancel}
\usepackage{ulem}
\usepackage{amsfonts}
\usepackage{amssymb}
\usepackage{graphicx}
\usepackage{amsmath}
\usepackage{enumerate}
\usepackage{mathtools}
\usepackage{subfig}
\usepackage{color}
\usepackage{float}
\usepackage{here}
\usepackage{cite}
\usepackage{mathrsfs}
\usepackage{float,epsfig}
\usepackage{dcolumn}

\usepackage{graphicx}
\usepackage{bm}
\usepackage{amsmath,amssymb,amsthm}
\usepackage[colorlinks=true,linkcolor=blue,citecolor=red]{hyperref}
\newcommand{\be}{\begin{equation}}
\newcommand{\ee}{\end{equation}}
\newcommand{\bea}{\setlength\arraycolsep{2pt} \begin{eqnarray}}
\newcommand{\eea}{\end{eqnarray}}

\def\0{{\sst{(0)}}}
\def\1{{\sst{(1)}}}
\def\2{{\sst{(2)}}}
\def\3{{\sst{(3)}}}
\def\4{{\sst{(4)}}}
\def\5{{\sst{(5)}}}
\def\6{{\sst{(6)}}}
\def\7{{\sst{(7)}}}
\def\8{{\sst{(8)}}}
\def\sst#1{{\scriptscriptstyle #1}}

\title{\bf \Large
 $f(R)$ Gravity Effects on  Charged Accelerating AdS Black Holes using  Holographic  Tools }

\author{
Adil  Belhaj$^{1}$\thanks{belhajadil@fsr.ac.ma}, Hasan  El Moumni$^{2}$\thanks{hasan.elmoumni@edu.uca.ma},  Karima  Masmar$^3$\footnote{karima.masmar@edu.uca.ma (Corresponding author)
} \footnote{ Authors in alphabetical order, they contributed equally to this work.}
\\
{\small $^{1}$ ESMAR, Physics Department, Faculty of Science,
Mohammed $V$ University in Rabat, Morocco. }\\
{\small $^{2}$ EPTHE, Physics Department, Faculty of Science,  Ibn Zohr University, Agadir, Morocco. }\\
  {\small $^{3}$ Laboratory of  High Energy Physics and Condensed Matter
HASSAN II University} \\{\small Faculty of Sciences Ain Chock, Casablanca, Morocco. }
}

\vspace{-3.6em}

\begin{document}
\maketitle
\begin{abstract}

We investigate numerically  $f(R)$ gravity effects  on  certain AdS/CFT tools  including  holographic entanglement entropy and two-point correlation functions  for a charged single accelerated Anti-de Sitter black hole in four dimensions. We find that
 both holographic entanglement entropy and two-point correlation functions decrease by  increasing acceleration parameter $A$,  matching perfectly with literature. Taking into account   the $f(R)$ gravity parameter $\eta$,  the decreasing scheme of the holographic quantities persist. However,   we observe  a  transition-like point where the behavior of the holographic tools change.  Two  regions meeting at  such a  transit-like point   are shown  up.  In  such a  nomination,   the first one is   associated with slow accelerating black holes while  the second  one corresponds to a fast accelerating  solution. In the first region,  the holographic entanglement entropy and two-point correlation functions decrease by increasing the $\eta$ parameter. However,   the behavioral  situation is  reversed  in the second one.
  Moreover, a cross-comparison between the entropy and the holographic entanglement entropy   is presented   providing  another counter-example showing  that such two  quantities do not  exhibit similar behaviors.

{\noindent}

\end{abstract}
\newpage
\tableofcontents


\section{Introduction}

Anti-de Sitter/Conformal field theory (AdS/CFT) conjecture has brought    new   roads  to understand  the quantum gravity \cite{Maldacena:1997re,Gubser:1998bc,Witten:1998qj,Aharony:1999ti}.  Motivated by the holographic entanglement entropy, an intrinsic potential interplay  between quantum information theory and  gravity  physics  has  been  established\cite{Ryu1,Ryu2}.

The entanglement entropy, which may be viewed as the von-Neumann entropy reflecting the entanglement between two different subsystems $A$ and $B$,  can be applied to the black hole. It has been  considered also as a candidate not only for  the statistical origin of the black hole entropy\cite{Strominger:1996sh}  but also for solving the information puzzle in black hole physics \cite{Strominger:1994sh}. It is worth noting that  the entanglement entropy acquires a similar geometric description to  the  bulk thermal entropy using   holography \cite{tHooft:1984kcu,Frolov:1993ym,Emparan:2006ni}.  Unfortunately, many  attempts which   have been made  to explain the BH entropy in terms    of  quantum entanglement suffer from   conceptual and technical difficulties.  Indeed, the  first difficulty is  associated with  the usual statistical interpretation of the black hole entropy in terms of microstates. This  is conceptually very different from the entanglement entropy measuring  the observer lack of information  by regarding the   system  as  quantum states  in an inaccessible region of the spacetime \cite{Cadoni:2010kla,Fursaev:2007sg}.  The second one  comes from  the fact that the entanglement entropy depends  on the number of species of matter fields. The associated  entanglement should reproduce the BH entropy. It depends also on the value of the ultraviolet (UV) cut-off which may be  needed to regularize the divergences  corresponding to  sharp boundaries  between the accessible and inaccessible regions of spacetime. However,  the Hawking-Bekenstein entropy should be universal.
In the context of the extended phase space,  it has been shown that the  entanglement entropy undergoes qualitatively a similar behavior as entropy  \cite{Johnson:2013dka}.  In parallelled ways,    many  investigations  have been devoted to such a  similitude \cite{Nguyen:2015wfa,Mo:2016ijb,Zeng:2015wtt,ElMoumni:2018fml}. However,   it has been revealed that  behavioral coincidence is associated with  the first law of black hole thermodynamics. This can also supported by the fact  the entanglement entropy depends on  the black hole mass  \cite{Sun:2016til,McCarthy:2017amh}. It urns out that the holographic entanglement entropy is not dual to the black hole  one.  It should not be expected that they  present  the same behavior as observed in \cite{Sun:2016til,McCarthy:2017amh}.

On the other hand, the inflation theory and dark energy  can be exploited to build Einstein modified gravity.  In this way,  the $f(R)$ gravity  can be considered a  good example to approach such  modified  theories.  Adding  $f(R)$  as  higher powers of $R$, one can  implement   the Ricci and Riemann tensors and their derivatives in Lagrangians, which could describe  possible physical models\cite{Upadhyay:2018ykz, Khurshudyan:2014iua, Capozziello:2017zow, Buchdahl:1983zz, Starobinsky:1980te, DeFelice:2010aj, Capozziello:2011et, Nojiri:2010wj, Sadeghi:2015nda,Bueno:2016ypa}. In this  context, considerable efforts have been made to investigate the AdS black hole thermodynamics in the background of $f(R)$ gravity with constant curvature \cite{Moon:2011hq,Chabab:2018zix,Belhaj:2019idh} as in the context of the  AdS/CFT conjecture \cite{Bazeia:2014xxa,Pourhasan:2014fba,Belhaj:2019idh}.
Recently, several   results  have  been elaborated and refined   to support such  investigation  lines including accelerating
black holes. The latters,     being  known to be described by the so-called
C-metric \cite{50,51,52,53},   have  been used  not only to  study  the
pair creation of black holes \cite{54}, the splitting of cosmic
strings \cite{55,56}, but also  to construct the black rings in five-dimensional gravity, motivated by string theory and related inspired models \cite{57}.
The thermodynamics of accelerating black holes \cite{58},  generalized
 results to the case of varying conical deficits for
C-metric,  and their applications to holographic heat engines,   have been studied in many places, see for instance \cite{58,59, Balasubramanian:1999zv,Zhang:2018vqs,Zhang:2018hms,Rostami:2019ivr}.

The aim of this paper is contribute to these activities by  investigating numerically  the  effect of  the $f(R)$ gravity   on  certain AdS/CFT tools  including  holographic entanglement entropy and two-point correlation functions  for a charged single accelerated Anti-de Sitter black hole in four dimensions.  It has been shown
that both holographic entanglement entropy and two-point correlation functions decrease by  increasing  the acceleration parameter $A$,  matching perfectly with literature \cite{Ref1,Ref2}. Taking into account   the $f(R)$ gravity parameter $\eta$,  the decreasing scheme of the holographic quantities persist. However,    it has been found   a critical-like point where the behavior of the holographic tools change.  Two  possible  regions  intersecting at  such a  transition-like point   are  observed.    The first one  corresponds to  slow accelerating black holes while  the second  one  is associated  with  a fast accelerating  solution. In the first region,  the holographic entanglement entropy and two-point correlation functions decrease by increasing the $\eta$ parameter. However,   the  second one corresponds to  a reversed   behavioral situation.

 The organization of the paper is as follows.  We  first  review, in section \textcolor{blue}{Sec.}\ref{sectw},  the essential of the solution describing
 charged single accelerated AdS Black holes in $f(R)$ gravity backgrounds and the associated relevant thermodynamics. In section \textcolor{blue}{Sec.}\ref{part3} and \textcolor{blue}{Sec.}\ref{part4},  we, numerically, study  the entanglement entropy and two-point correlation functions and we prob the effect of the $f(R)$ gravity parameter $\eta$ on such holographic quantities. The last section is devoted to conclusion and  some open questions.

\section{Charged accelerating $f(R)$ black hole background}\label{sectw}
To start,  one  reconsiders the study of physical content of  $f(R)$ gravity. According to \cite{mannfr},  its   action   containing  a Maxwell gauge  term takes the following general form
\begin{equation}\label{sou}
\mathcal{S}=\frac{1}{16\pi G}\int d^{4}x\sqrt{-g}{\cal{L}}.
\end{equation}
In this action,  $\cal{L}$  can be expressed as
\begin{equation}\label{wswsws}
{\cal{L}}=R+f(R)-F_{ab}F^{ab},
\end{equation}
  where $R$ is the Ricci scalar.   It is noted that  $f(R)$ is an auxiliary function of $R$, which determines the gravity model \cite{Hod:2000kb,Nojiri:2006ri}.
  The  field strength $F_{ab}=\nabla_{a}B_{b}-\nabla_{b}B_{a}$ is  derived  from the gauge potential one form  $B_{a}$. Using field theory technics  associated with the variation
with respect to the metric,  and the gauge field,  one can get  field equations of motion.  Roughly, they are listed as follows
\begin{eqnarray}\label{eos1}
R_{ab}&-&\frac{1}{2}Rg_{ab}-\frac{1}{2}f(R)g_{ab}+R_{ab}f^{\prime}(R)-\nabla_b \nabla_a R f^{\prime\prime}(R)+g_{ab}\nabla_{c}\nabla^{c}Rf^{\prime\prime}(R)\nonumber\nonumber\\&-&\nabla_a R \nabla_b R f^{(3)}(R)+ g_{ab}\nabla_c R\nabla^{c} Rf^{(3)}(R)=2T_{ab}
\end{eqnarray}
and
\be\label{eos2}
\nabla_b \nabla^b B^a-\nabla_b \nabla^a B^b=0,
\ee
where one has used
\begin{eqnarray}
f^{\prime}(R)&=&\frac{df(R)}{dR}\\
T_{ab}&=&F_{a}^{~c}F_{bc}-\frac{1}{4}F_{cd}F^{cd}g_{ab}.
\end{eqnarray}
One can  check that the electromagnetic field $T_{ab}$ is traceless
\begin{equation}
T^{a}_{~a}=0.
\end{equation}
It turns out that a  maximally symmetry solution  can be obtained   by considering a  constant  Ricci scalar \cite{Sotiriou:2008rp}.  In the  situation given  by $R=R_0~(R_{0}\neq 0)$,  the metric   field equation  reduces to \cite{Moon:2011hq,Sheykhi:2012zz}
\be\label{sef}
R_0 -R_{0} f^{\prime}(R_0) +2 f(R_0)=0.
\ee

  In this way,   Eq.\eqref{eos1} can be rewritten  as follows
\be\label{eos11}
\eta R_{ab}-\frac{\eta}{4}R_{0}g_{ab}=2T_{ab}.
\ee
Here,  $\eta$  is defined as
\begin{equation}\label{fr2}
\eta =1+f^{\prime} (R_0).
\end{equation}

Exploiting  the results  reported in \cite{mannfr} and using the    equation of motion given by Eq.\eqref{eos11}, we can  derive  a charged accelerating black hole solution based on  the following
line element
\begin{eqnarray}\label{met}
\begin{aligned}
ds^2=\frac{1}{\Omega^2}&\left[-\frac{N(r)dt^2}{\alpha^2}+\frac{dr^2}{N(r)}
+r^2 \left(\frac{d\theta^2}{g(\theta)}+g(\theta)\sin^2\theta \frac{d\phi^2}{K^2}\right)\right].
\end{aligned}
\end{eqnarray}
The terms  involving such a line element  are given by
\begin{eqnarray}
\Omega&=&1+Ar\cos\theta,\\
N(r)&=&(1-A^2 r^2)\left(1-\frac{2m}{r}+\frac{q^2}{\eta r^2}\right)-\frac{R_0 r^2}{12},\\
g(\theta)&=&1+2mA\cos\theta+\frac{q^2}{\eta}A^2\cos^2 \theta,\\\label{alpha}
\alpha&=&\sqrt{\Xi (1+\frac{12 A^2 \Xi}{R_0})},\\
\Xi&=&1+\frac{q^2 A^2}{\eta}.
\end{eqnarray}
In such a  black hole solution, the conformal factor $\Omega$  denotes  the conformal boundary $r_{b}$ of the black hole given by  $r_{b}=-1/(A\cos\theta)$. $A,~m$ and $~q$ indicate  individually the acceleration, the mass parameter and the electric charge parameter of the black hole, respectively. $K$ is  the conical deficits of the black hole on the north and south poles \footnote{The present space-time involves conical singularities  localized  at $(\theta=0,\pi)$. In this way,  the  metric regularity condition  at the poles, $\theta_+=0$ and $\theta_-=\pi$ imposes  that
\begin{equation}
  K_{\pm}=g(\theta_{\pm})=1+2 m A+\frac{q^2 }{\eta}A^2.
\end{equation}
Generally, $K$ is  explored  to regularize one pole  producing a  conical deficit or a conical excess along the other pole.  Due to   a negative energy  corresponding to   the source of a conical excess, it  has been supposed, throughout this  study,  that the black hole is regular on the north pole where $\theta=0$ and $K =K_+$.  It turns out that   a conical deficit can  exist on the other pole for                                                                                               $\theta=\pi$. }.   In order to  get a normalized Killing vector at the conformal infinity \cite{Podolsky:2002nk,Anabalon:2018ydc,Anabalon:2018qfv}, the parameter $\alpha$  can be  exploited  to rescale the time coordinate.   When   $A$  goes to zero,  it is  observed  that  one   can  recover the usual  charged AdS black hole in the $f(R)$ gravity\cite{Nojiri:2014jqa}.

It is noted that $R_{0}<0,~R_{0}=0$ and $R_{0}>0$  are associated with asymptotically AdS, flat, and dS accelerating black holes, respectively. In the  present investigation,  we will  be only interested in  the case $R_{0}<0$.  For  $A=0,~K=1$ and $~R_{0}<0$, the solution  reduces to the $f(R)$ black hole \cite{Moon:2011hq}. In this way,  the condition $\eta>0$ is needed to  insure  the existence of inner and outer horizons  \cite{Sheykhi:2012zz}. For certain physical  reasons, we will only deal with  the case $\eta>0$ for the accelerating $f(R)$ AdS black holes.  We hope,  other non trivial cases could be deal with in future works. Solving the equation of motion (\ref{eos2}) for the gauge field $B_a$, one  can obtain the electromagnetic tensor
\begin{equation}
F_{ab}=(dB)_{ab}.
\end{equation}
The corresponding  calculation gives
\begin{equation}\label{B-pot}
B_{a}=\frac{1}{\alpha}\left(\frac{q}{r_+}-\frac{q}{r}\right)(dt)_{a},
\end{equation}
where  $r_{+}$\footnote{The complete form of the $r_+$ is given in the appendix.} is the outer horizon. It is worth noting that  for
\begin{equation}
f^{\prime}(R_0)=0,~R_0 =-\frac{12}{\ell^2},
\end{equation}
the solution coincides with the charged accelerating AdS black hole in  the Einstein gravity framework.  For   $R_0= -12/\ell^2$,  the blackening factor of the black hole can be written as
\begin{equation}\label{mmet}
N(r)=\left(1-\frac{2m}{r}+\frac{q^2}{\eta r^2}\right)(1-A^2 r^2)+\frac{r^2}{\ell^{2}}
\end{equation}
where   $\eta$ and $\ell$ are considered as  independent  physical parameters.

\section{Holographic entanglement entropy of charged accelerating black holes in $f(R)$ gravity background} \label{part3}
\begin{figure}[!ht]
		\begin{center}
		\centering
			\begin{tabbing}
			\centering
			\hspace{9.3cm}\=\kill
			\includegraphics[scale=.58]{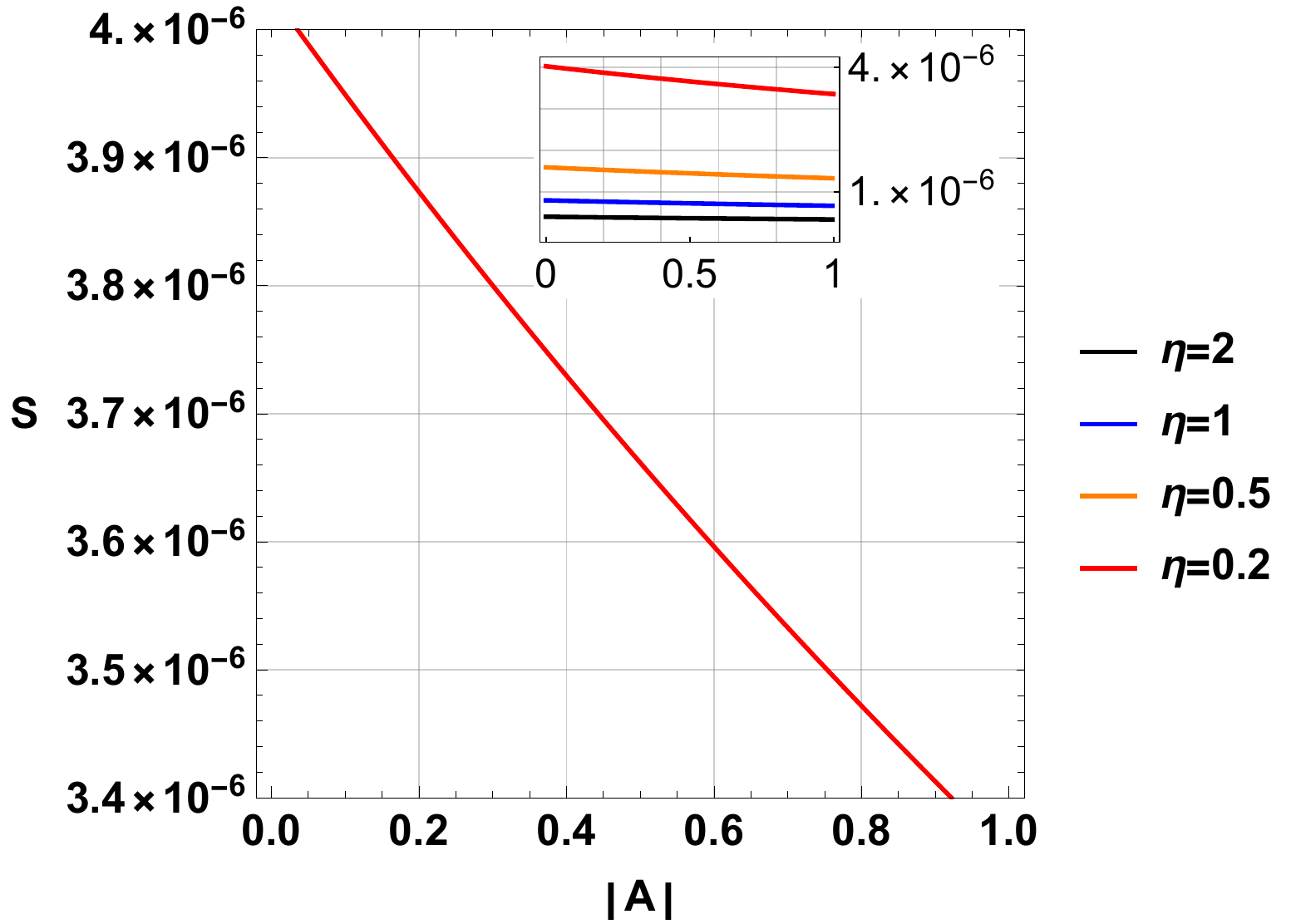} \>
			\includegraphics[scale=.5]{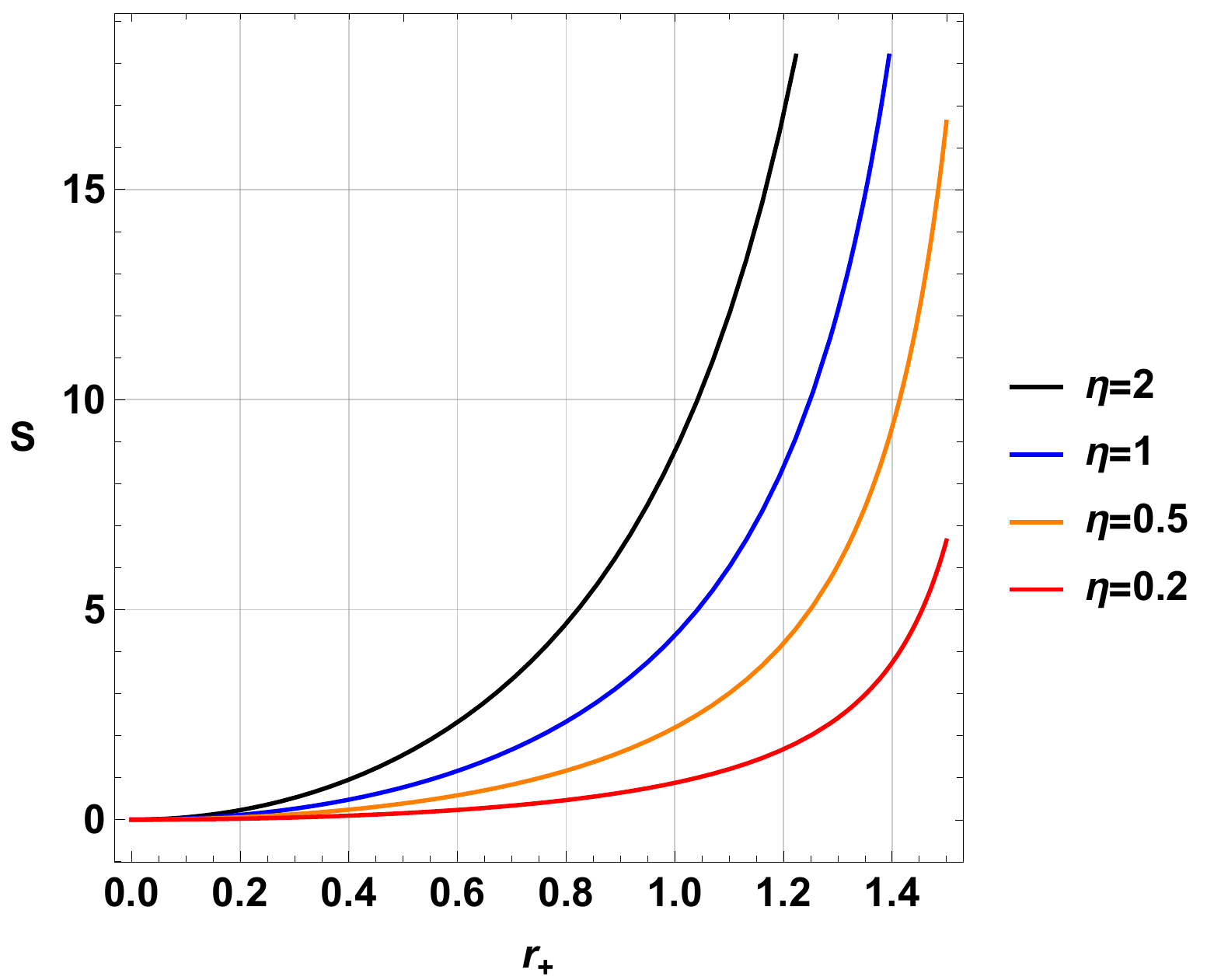} \\
		   \end{tabbing}
\caption{\footnotesize {\bf Left:} The
  entropy of the charged accelerating AdS Black hole with respect to the acceleration parameter $A$. Here, we take $q=0.01$, $m=0.1$, $\ell=1$,  $\eta=0.2$ and $r_+=0.190382$ given by Eq.\eqref{rhhh}. {\bf Right:} The
  entropy of the charged accelerating AdS Black hole in terms of  the horizon radius $r_+$ within different values of accelerating parameter $A$. Here, we take $q=0.01$, $m=0.1$ and $\ell=1$,  }\label{figure311}
\end{center}
\end{figure}

In the present investigation, we ignore the acceleration horizons of  C-metrics  and consider only the black hole horizon  in order to  get a  well-defined temperature.
This  kind of simplification is known as the "slowly accelerating C-metric",   proposed in \cite{Podolsky2}. It has been remarked that various solutions can appear depending on  the parameter $A$. For $A < 1/\ell$,  it has been  suggested that a single black hole can be built  in the AdS  geometry space with the only horizon being that of such a  black hole.     For $A > 1/\ell$,  however,  it is involved  two black holes separated
by the acceleration horizon \cite{Dias,Podolsky2,Krtous}.  According to \cite{Hong:2003gx}, a C-metric with a cosmological constant and charge in the Hong-Teo coordinates can be  represented by the  metric given in  Eq.\eqref{met}. The  first law of thermodynamics for accelerating black holes with a varying conical deficit and a critical behavior  has been investigated  in many places including \cite{Kubiznak1}.  It  has been  extended to  $f(R)$ gravity backgrounds in \cite{mannfr}.  The mass $M$ of the slowly accelerating $f(R)$ AdS black hole can be calculated as
\begin{equation}\label{mass}
M=
\frac{\eta m(1-A^2 l^2 \Xi)}{K\alpha}.
\end{equation}
While its charge is given by
\begin{equation}\label{ele}
Q=\frac{1}{4\pi }\lim_{\Omega\to 0}\int\ast F=\frac{1}{4\pi}\int \frac{q}{K}\sin\theta d\theta d\phi =\frac{q}{ K}.
\end{equation}
 Roughly,  the entropy of the black hole   reads as
\begin{equation}\label{accent}
S=-2\pi\oint d^{2}x\sqrt{\hat{h}}\frac{\partial \mathcal{L}}{\partial R_{abcd}}{\epsilon}_{ab}{\epsilon}_{cd}=\frac{\eta\pi r_{+}^{2}}{K(1-A^2 r_+^2 )},
\end{equation}

where $\hat{h}$ indicates  the determinant of the induced metric on the $t=\text{const.}$ and $r=r_{+}$ hypersurface. The quantity  $\epsilon_{ab}$ is a normal tensor   verifying   $\epsilon_{ab}\epsilon^{ab}=-2$.  In Fig.\ref{figure311}, we illustrate  the variation of such an entropy  quantity in terms of the accelerating parameter $A$ and the horizon radius $r_+$.


  The left side   of Fig.\ref{figure311},  associated with the  fixed horizon radius, shows  that the entropy presents a decreasing variation in terms of the accelerating parameter $A$ and the $f(R)$ gravity parameter $\eta$. For fixed  acceleration parameter $A$ (right side  of Fig.\ref{figure311}), it follows   that the entropy grows up within the horizon radius $r_+$. Moreover,
 we observe  that entropy increases by  increasing $ r_+$ and blows up at $(Ar_+)^2=1$. Following $\eta$ parameter,  the entropy grows as $\eta$ increases.

Roughly, the temperature of the black hole is  given by
\begin{eqnarray}\label{temp}
T&=&\frac{N^{\prime}(r_+)}{4\pi \alpha}\nonumber\\&=&\frac{ \ell^2 \left(A^2 r_+^2-1\right)^2 \left(q^2-\eta  r_+^2\right)-\eta  r_+^4 \left(3-A^2 r_+^2\right)}{4 \ell^2 \pi  \alpha  \eta  r_+^3 \left(A^2 r_+^2-1\right)}\nonumber \\&=&\frac{r_+}{2 \pi  \alpha  \ell^2 \left(1-A^2 r_+^2\right)}+\frac{\left(A^2 r_+^2-1\right) \left(q^2-\eta  m r_+\right)}{2 \pi  \alpha  \eta  r_+^3}.
\end{eqnarray}

It  has been  quite well-known  that  the  entanglement entropy (EE) is considered  as  a good  approach  to measure the amount of quantum information of a bipartite system. One way to quantify this information is to compute  the von Neumann entropy of a bipartite system where the system is divided into two parts. For this proposal,
we first consider a time slice located at the AdS spacetime. Then,  we
 calculate  the area of the minimal surface $\gamma_A$. This can be  parameterized by $r=r(\theta)$ given by a time slice $t= 0$ in the line-element appearing in  Eq.\eqref{met}.  In this case,  the entanglement entropy can be expressed as
 \begin{equation}\label{En}
 S_A=\frac{Area(\gamma_A)}{4 G}=\frac{\mathcal{A}}{4 G}\ ,
 \end{equation}
where $G$ is the Newton's constant \cite{Ryu1,Ryu2}. Using Eq.(\ref{En}) and the metric presented in Eq.(\ref{met}), the holographic entanglement entropy (HEE) for charged accelerated black holes in $f(R)$ gravity can be examined.   The area, associated with the  minimal hyper-surface in the presence of a charged accelerating AdS black hole in the bulk, is computed using the  following relation
\begin{equation}\label{A}
{\cal{A}}= 2\pi\int_0^{\theta_0}\frac{r(\theta) \sin\theta \ \sqrt{g(\theta)}}{\Omega ^2 K} \sqrt{\frac{r'(\theta)^2}{N(r(\theta))} + \frac{r(\theta)^2}{g(\theta)}}d\theta\,.
\end{equation}
In fact, this surface  area should be minimized.  A Lagrangian, associated with  such an  area,   can be worked out to give   the  equation of  motion. Indeed, the latter reads as
\begin{eqnarray}
\label{rth2}\nonumber
&&g(\theta )^2 \Omega  \sin \theta  N'(r(\theta) ) r'(\theta )^3+2 N(r(\theta) )^2 r(\theta ) \Big[r'(\theta ) \Big(2 g(\theta )
\sin \theta  r(\theta ) \Omega '-\Omega  \big[\sin \theta  r(\theta ) g'(\theta )\\&&+g(\theta ) (\cos \theta  r(\theta )-3 \sin \theta
r'(\theta ) )\big]\Big)-g(\theta ) \Omega  \sin \theta  r(\theta ) r''(\theta )\Big]+4 N(r(\theta) )^3 \Omega  \sin \theta  r(\theta )^3
\\
&&\nonumber N(r(\theta) ) g(\theta ) r'(\theta ) \Big(2 \Omega  \sin \theta  r(\theta )^2 N'(r(\theta ))-r'(\theta )^2 [\Omega  \sin \theta
 g'(\theta )-4 g(\theta ) \sin \theta  \Omega '+2 g(\theta ) \Omega  \cos \theta ]\Big)=0\,.
\end{eqnarray}
A priori, they are  many ways to  solve this equation. However, the analytic one seems to be hard. For such a reason,   we  could   use  numerical procedures  to get  $r(\theta)$\footnote{The Wolfram Mathematica program  used to support the findings of this study is available from the corresponding author upon request}.

\begin{figure}[!ht]
  \center
  \includegraphics[scale=1]{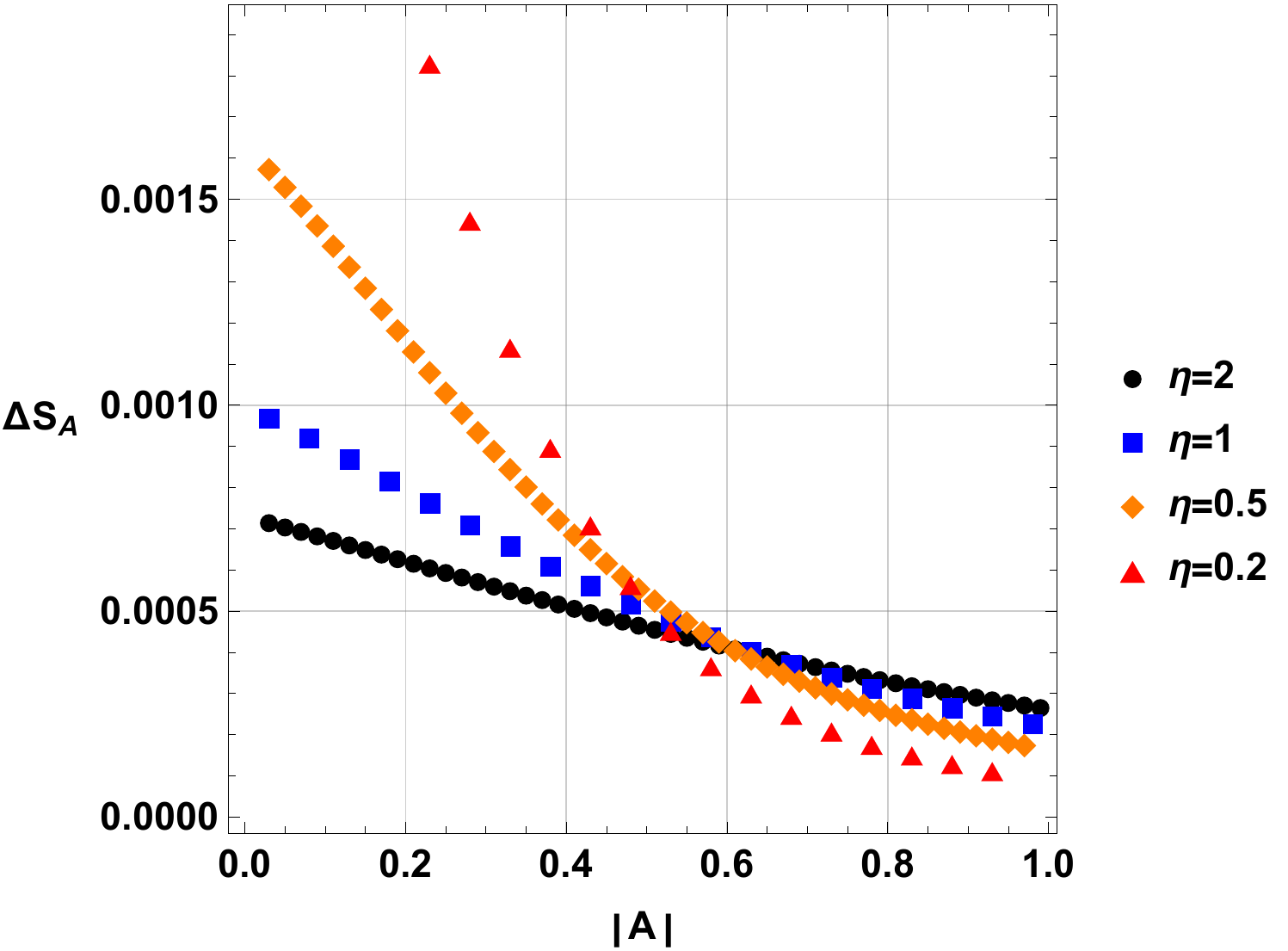}\\
  \caption {\it{ \footnotesize The holographic entanglement entropy for the charged accelerating AdS Black hole with respect to the acceleration parameter $A$. Here, we take $q=0.01$, $m=0.1$, $\ell=1$ and $\theta_0=0.16$. The holographic entanglement entropy
   decreases
with increasing  acceleration  exhibiting   behaviour similar to the one observed in relativistic quantum information studies \cite{Ref1,Ref2}.}}
\label{figure1}
\end{figure}

In this way,  the existence of the conformal factor $\Omega$  changes significantly  the location of the AdS boundary from infinity $r = \infty$ to finite values $r =-1/ A \cos\theta$ associated with vanishing $\Omega$. Then, now all  the boundary conditions are $r'(\theta)=0$ and $r=r_0$ at $\theta=0$ and $r=-1/(A \cos \theta_0)$ at $\theta=\theta_0$.
To regularize the entanglement entropy, we follow the result of \cite{NPB,Xu}. Indeed,    we subtract the area of the minimal surface in  Rindler-AdS whose boundary is also $ \theta=\theta_0$ with 
\begin{eqnarray}\label{rrads}
&&r_{RAdS}(\theta)=\frac{\ell}{\sqrt{\frac{\cos ^2 \theta }{\cos ^2 \theta _0}-1}}+\frac{\ell^2 \cos \theta }{\cos ^2 \theta-\cos ^2 \theta _0} A+\frac{ \ell \cos \theta _0}{4 (\cos ^2 \theta -\cos ^2 \theta _0 )^{3/2}}\times
\nonumber\\
&&\Big[(4 \ell^2+1) (2 \cos ^2\theta -1)-(2 \cos ^2\theta _0-1) (2 \cos ^2 \theta +2 \ell^2)+6 \ell^2+1\Big] A^2
+O(A^3)\,.
\end{eqnarray}
Substituting $r(\theta)$ from  Eq.(\ref{rth2}) in Eq.(\ref{A}) and with the help of Eq.\eqref{rrads}, we can illustrate  the holographic entanglement entropy with respect to the acceleration parameter $A$ as shown in Fig.  \ref{figure1},
where $e=0.01$, $m=0.1$,  $\ell=1$ and $\theta_0=0.16$. The ultra-violet cutoff is chosen to be $\theta_c=0.1599$.

It follows that  the values of $\Delta S_A$ decrease as a function of the acceleration parameter when the charged accelerating AdS black hole is located in the bulk. At this level,  one may give some  discussing points. The first one concerns the similarity with the behaviors reported in the study of the entanglement entropy within the framework of the relativistic quantum information theory \cite{Ref1,Ref2}. The second point  is that the variation of such a  quantity is affected by the values of $\eta$. Two regions associated with such behaviors can appear. For the small values of the accelerating parameter $A$,   one can note that the entanglement entropy
 is large as the value of $\eta$ is small. While in the large values of $A$,  the behavior changes   contrarily with respect to  the first region.  These regions meet at  a  transition-like point,  where all the graphs intersect.

 Before going ahead, a brief comment on the behavior of the entropy given by Eq.\eqref{accent} and the holographic entanglement entropy given by Eq.\eqref{En} is needed.  Indeed, both the quantities share the same behavior apparently.  Indeed, they both   decrease when the accelerating parameter $A$ increase. For  the $\eta$ parameter, however,  the transition point observed in the holographic picture is absent in the entropy one.
Since   the holographic  entanglement entropy is not dual to the black hole entropy, they should not necessary  present the same behavior.  This difference observed here commensurate with
 \cite{Sun:2016til,McCarthy:2017amh} where the authors demonstrate that the Maxwell law observed in $(T,S)$ plane is not verified in $(T,\Delta S_A)$ one.  The apparently similitude,   observed in this case,  is due to the fact that HEE depends on mass directly.

\section{Two-point correlation function of charged accelerating black holes in $f(R)$ gravity background}\label{part4}

Having  investigated  the behavior of the holographic entanglement entropy
as  a function of the accelerating parameter $A$ and   the parameter $\eta$ , we move to discuss
 two point correlation functions. In particular, we will   show that  such a local  observable exhibits the same scheme found   in the previous section. To show that,  we first  give a concise review  of the Maldacena correspondence  in order to introduce
the time two point correlation functions  under the saddle-point approximation and in the large limit of $\Delta$ as \cite{Balasubramanian:1999zv}
\begin{equation}
\langle {\cal{O}} (t_0,x_i) {\cal{O}}(t_0, x_j)\rangle  \approx
e^{-\Delta {L}}. \label{llll}
 \end{equation}
It is noted that  $\Delta$ stands for the conformal dimension of the scalar operator $\mathcal{O}$ in the dual field theory.   $L$ is the length of the bulk geodesic between the two  points $(t_0, x_i)$ and $(t_0, x_j)$ on the AdS boundary.
Using  spacetime symmetry arguments of the  associated  black hole,
  we can redefine   $x_i$ as $\theta$ with the boundary $\theta_0$   to  parameterize the trajectory. In this way,  the proper length   takes the following form
\begin{eqnarray}
L=\int_0 ^{\theta_0}\mathcal{L}(r(\theta),\theta) d\theta,~~\mathcal{L}= \frac{1}{\Omega ^2 } \sqrt{\frac{r'(\theta)^2}{N(r(\theta))} + \frac{r(\theta)^2}{g(\theta)}}\,,
 \end{eqnarray}
where  one has used $r^{\prime}=dr/ d\theta$. Treating  $\mathcal{L}$ as Lagrangian
 and $\theta$ as time,  the equation of motion  associated with $r(\theta)$  reads as
\begin{eqnarray}
& &
4 r(\theta )^3 N(r(\theta ))^2 \Omega '(r(\theta ))-4 g(\theta ) N(r(\theta )) \Omega (r(\theta )) r'(\theta )^2-2 r(\theta )^2
   N(r(\theta ))^2 \Omega (r(\theta ))\nonumber\\
   & &+ r(\theta ) \left[r'(\theta )^2 \left(4 g(\theta ) N(r(\theta )) \Omega '(r(\theta
   ))-g(\theta ) \Omega (r(\theta )) N'(r(\theta ))\right)+N(r(\theta ))
   g'(\theta ) \Omega (r(\theta )) r'(\theta )\right. \nonumber\\ & &\left.+2 g(\theta ) N(r(\theta ))
   \Omega (r(\theta )) r''(\theta )\right]=0.
\end{eqnarray}

Using  the same boundary conditions used in he  previous section
$r'(\theta)=0$ and $r=r_0$ at $\theta=0$ and $r=-1/(A \cos \theta_0)$ at $\theta=\theta_0$, we
 can solve such an equation  by  choosing the same values of   $\theta_0$ with the  same UV cutoff
in the dual field theory.  Labeling  the regularized two-point correlation functions as $\Delta L_A=L-L_0$,  where  $L_0$ is the {geodesic length in  Rindler-AdS under the same boundary region obtained  with the help of Eq.\eqref{rrads},   %
we present  the variation of the function of $\Delta L_A$ in terms of the acceleration parameter $A$. The associated computation is illustrated in  Fig.\ref{figure3}. It is observed, from this figure,   that  the same behavior is held as  the HEE picture.
 \begin{figure}[!ht]
  \center
  \includegraphics[scale=1]{{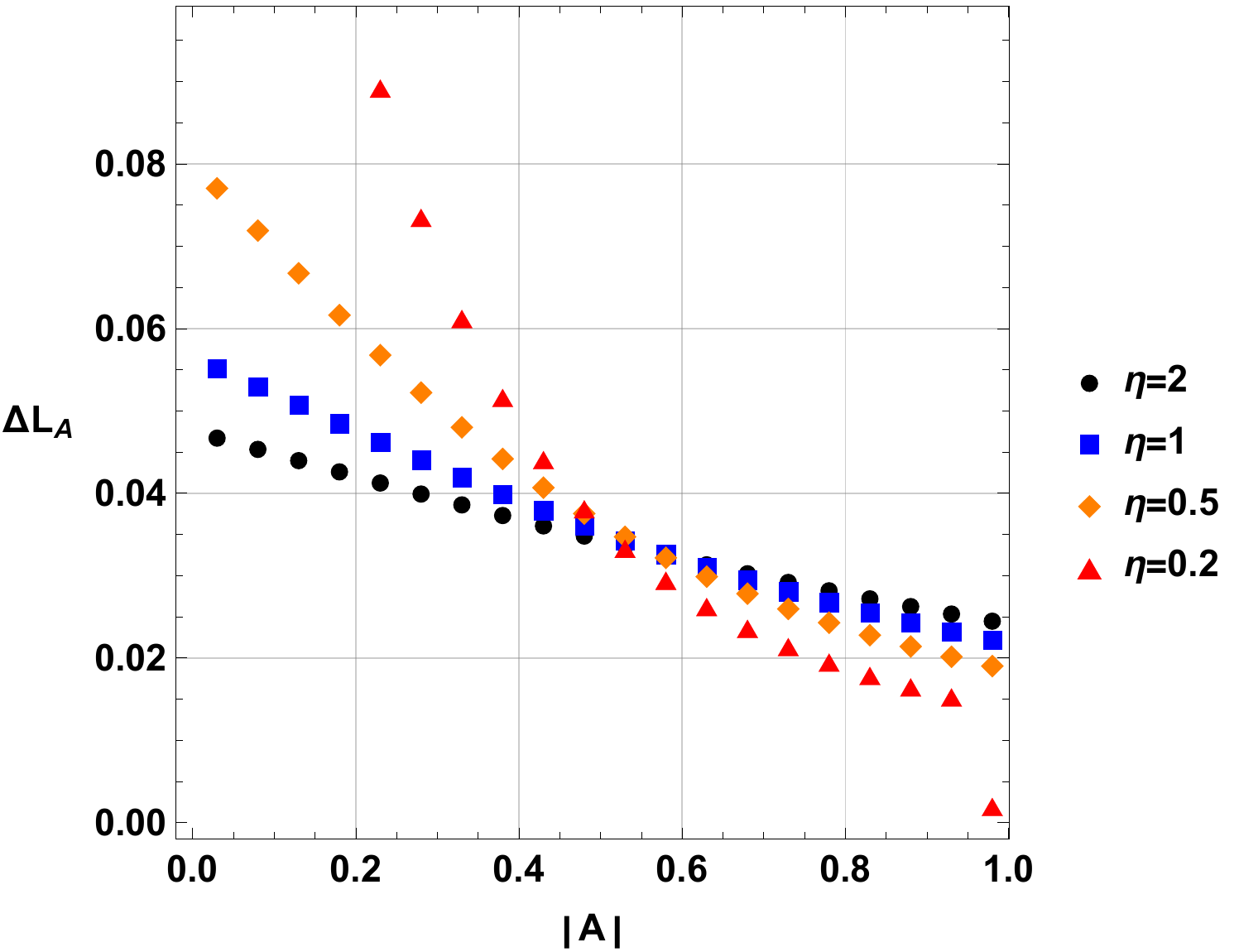}}\\
  \caption {\it{ \footnotesize
The  two-point correlation functions for a charged accelerating AdS Black hole with respect to the acceleration parameter $A$. Here, we take $q=0.01$, $m=0.1$, $\ell=1$,  $\theta_0=0.16$ and $\theta_c=0.1599$.}}
\label{figure3}
\end{figure}

\newpage

\section{ Discussions and  concluding remarks}
  In this  paper,  the holographic tools  including  entanglement entropy and two-point correlation functions   have been investigated  for accelerating observers in  four dimensional $f(R)$  gravity backgrounds. In particular, we have  considered  a C-metric with  a cosmological constant in the Hong-Teo coordinates to evaluate  the minimal surface area associated with  HEE   and two point correlation functions  of  a  charged single accelerated AdS black hole in four dimensions.

Concretely,  we  have dealt with   the  acceleration parameter less than the inverse of the cosmological length $\ell$, where a single black hole  appears in  the AdS geometry  with the only  horizon being that of  a  black hole.
 Due to the appearance of a communicating  horizon, there is no possibility for a uniformly accelerated observer to get  information about space-time behaviours.  A loss of information  takes place  generating an   entanglement degradation which implies that, in a space-time with horizons, one might expect some kinds of  information loss puzzle \cite{Polchinski:1994zs,Hod:2000kb}. For  such a reason,  various activities   have been made  to figure out how the holographic entanglement entropy and two point correlation functions behave as  a function of the accelerating  parameter $A$ and the parameter $\eta$ controlling  $f(R)$ gravity effects.  It has been observed that  the holographic entanglement entropy
and two point correlation  functions   decrease  by  increasing acceleration parameters  of such  black hole   solutions. This indicates that the holographic entanglement entropy follows the universal behavior of entanglement entropy for accelerating observers \cite{Ref1,Ref2}.

Moreover, we have investigated  effects of the $f(R)$ gravity parameter $\eta$ on such nonlocal observers. Rigorously,  the  holographic quantities including HEE  and  two point correlation functions exhibit non trivial behaviors.  It has been
 observed   a transition point where the behavior of the holographic tools change.  We have found  two  regions  intersecting  at such a  transition point.    These two regions correspond to slow and fast  accelerating   black holes  in our nomination,  retrospectively. In the first region,  HEE and two-point correlation functions decrease by increasing the $\eta$ parameter. However,    the second one is associated with a reversed  behavioral  situation.  A  comparison between the non local observables  and the entropy have been investigated. The absence of the transition point in the entropy picture provides   a counter example showing that  both entropy and nonlocal observable exhibit a similar behavior.

This work comes up with some  open questions.  It  will be interesting to extend  the calculation for  $A > 1/\ell$ with two black holes separated by an acceleration horizon. Motivated by string theory and related topics, higher dimensional solutions could be developed  and investigated in the presence of non trivial contributions including dark energy effects.  Moreover,  the thermodynamical behavior within the nonlocal observable quantities  could be considered  as  interesting  investigated  approaches.
We leave these questions for future works.

\section*{Data Availability}
Since this work is theoretical, there is no data used to support the findings of this study. 
The Mathematica notebooks associated with this study are available from the corresponding author upon request.\section*{Conflicts of Interest}
The author declares that they have no conflicts of interest.

\section*{Acknowledgements}
 This work is partially supported by the ICTP through AF-13.   We are grateful to the anonymous referee for  his careful reading of our manuscript,  insightful comments and suggestions, which have allowed us to improve this paper significantly.
 A special thanks to Pr. Ming Zhang for the interesting discussion about the parameter $\eta$ and the $f(R)$ gravity model and Pr. Masoumeh Tavakoli  for useful correspondence.

\section*{Appendix}
\begin{appendix}
The complete form of the event horizon radius is
\begin{equation}\label{rhhh}
r_+=\frac{1}{2} \sqrt{\frac{3 a_2^2-8 a_1 a_3}{12 a_1^2}+X+Y}+\frac{1}{2} \sqrt{\frac{a_2^2}{2 a_1^2}-\frac{8
   a_1^2 a_4-4 a_1 a_2 a_3+a_2^3}{4 a_1^3 \sqrt{\frac{3 a_2^2-8 a_1 a_3}{12
   a_1^2}+X+Y}}-\frac{4 a_3}{3 a_1}-X-Y}-\frac{a_2}{4 a_1},
\end{equation}
where
\begin{eqnarray}\nonumber
X&=& \frac{\sqrt[3]{2} \left(12 a_1 a_5-3 a_2 a_4+a_3^2\right)}{3 a_1 Z},\\ \nonumber
Y&=& \frac{Z}{3 \sqrt[3]{2} a_1},\\ \nonumber
Z&=&\bigg(\sqrt{\left(-72 a_1 a_3 a_5+27 a_1 a_4^2+27 a_2^2 a_5-9 a_2 a_3 a_4+2
   a_3^3\right)^2-4 \left(12 a_1 a_5-3 a_2 a_4+a_3^2\right)^3}\\&-&72 a_1 a_3 a_5+27 a_1
   a_4^2+27 a_2^2 a_5-9 a_2 a_3 a_4+2 a_3^3\bigg)^{\frac{1}{3}},
\end{eqnarray}
and
\begin{eqnarray}\nonumber
a_1&=& \eta -A^2 \eta  \ell ^2     ,\\ \nonumber
a_2&=&  2 A^2 \eta  m \ell ^2    ,\\ \nonumber
a_3&=&   \eta  \ell ^2-A^2 q^2 \ell ^2   ,\\ \nonumber
a_4&=&    -2 \eta  m \ell ^2  ,\\ \nonumber
a_5&=&     q^2 \ell ^2 .\\
\end{eqnarray}
\end{appendix}


\bibliographystyle{unsrt}
\bibliography{biblio.bib}

\end{document}